\documentstyle[12pt]{article}
\title{Illusory Quantum Hair}
\author{Aharon Casher \and Oskar Pelc}
\def\al{\alpha}
\def\bt{\beta}
\def\gm{\gamma}                
\def\dl{\delta}                \def\Dl{\Delta}

\def\kp{\kappa}
\def\lm{\lambda}               
\def\vph{\varphi}
\def\om{\omega}               \def\Om{\Omega}
\def\Ac{\mbox{\protect$\cal A$}}
\def\Dc{\mbox{\protect$\cal D$}}
\def\Oc{\mbox{\protect$\cal O$}}
\def\Pc{\mbox{\protect$\cal P$}}
\def\Zb{{\bf Z}}
\def\RR{I\!\!R}

\def\pt{\partial}
\def\goto{\rightarrow}

\def\derp#1{\frac{\pt}{\pt#1}}
\def\ie{{\em i.e.\ }}
\def\eg{{\em e.g.\ }}

\def\beq{\begin{equation}}
\def\eeq{\end{equation}}
\begin{document}
 
\input epsf
 
%\maketitle
\begin{titlepage}
 
\begin{flushright}
TAUP 2091-93\\
August 1993\\[5mm]
%\today  \\[5mm]
hep-th/9308107\\[20mm]
\end{flushright}
 
\begin{center}
\Large Illusory Quantum Hair \\[15mm]
\large Aharon Casher \hspace{5mm}\normalsize and
                     \hspace{5mm}\large Oskar Pelc\footnote{
 E-mail: casher@taunivm, oskar@taunivm}  \\[10mm]
\normalsize
{\em School of physics and astronomy \\
 Raymond and Beverly Sackler Faculty of Exact Sciences \\
 Tel-Aviv University, Ramat-Aviv, Tel-Aviv 69978, Israel}
\\[20mm]
\end{center}
 
\begin{abstract}
 We analyze the quantum hair model proposed recently by Coleman,
 Preskill and Wilczek. We give arguments suggesting that the
 potential hair is expected to be destroyed by the instability of
 the black-hole. We also discuss the general implications of such
 arguments on the prospects for formulating a quantum extension
 of the classical ``no hair'' theorems.
\end{abstract} 

\end{titlepage}
 
\section{Introduction}
The classical ``no hair'' theorems state that a stationary black hole
is characterized by a very limited set of parameters, essentially its
mass, angular momentum and charges corresponding to massless gauge
fields\footnote{for details, see \eg \cite[p. 322 -- 324]{Wald}}. This
means that an external observer  can get very little information about
the matter configuration that formed the black hole. This situation is
directly connected to the thermal nature of the Hawking radiation and
if the black hole evaporates completely this leads to a violent
contradiction with the unitary evolution of a system, according to
quantum theory.
 
One way out would be to find additional degrees of freedom of a black
hole that are observable only by quantum effects -- ``quantum hair''.
Coleman, Preskill and Wilczek (CPW) \cite{CPW} proposed recently a
model in which such a quantum hair seems to exist. Using
semiclassical thermodynamics, neglecting the black-hole evaporation
through the Hawking radiation, they calculated the effects of such a
hair that can be measured by an external observer.
 
The effect found is very small -- non-perturbative in $\hbar$ -- and
therefore a natural question that arises is whether it is reasonable
to assume that this effect is not influenced by the Hawking radiation
(which is of first order in $\hbar$). Our analysis indicates that it
is not. Indeed we show that the characteristic time of the effect is
much longer than the whole (semiclassical) life time of the black
hole and for such a long time the thermodynamic equilibrium hypothesis
is certainly not justified. This means that the results indicating the
existence of the quantum hair are not reliable. Furthermore, an
analysis of an analogous toy model suggests that the instability of
the black hole -- which causes the energy levels to have finite width
-- ``hides'' the potential hair.
 
We start, in section 2 by reviewing CPW's model, in section 3 we show
its inconsistency and in section 4 we analyze a toy model which
demonstrates what we expect really happens. We conclude by discussing
the general question of the existence of a quantum hair in view of the
above results.
 
\section{The Proposed Hair}
In this section we will review briefly the construction of the hair
proposed by CPW \cite{CPW}. The proposed model is an Abelian Higgs
model coupled to gravity, where the charge of the scalar field $\vph$
that ``condenses'' (acquires a non-trivial vacuum expectation value
$|<\vph>|=\frac{1}{\sqrt{2}}v$) is a multiple of the elementary charge
quantum: $Q_{\vph}=N\hbar e$. The action is (c=1)
\beq S=S_{grav}+S_{em}+S_{Higgs}+S_{mat} \eeq
where
\beq S_{grav}=-\frac{1}{16\pi G}
  \left[\int d^{4}x\sqrt{g}R+2\oint d^{3}x\sqrt{h}K\right]
\eeq
($h$ is the boundary metric and $K$ is the extrinsic curvature on the
boundary, both induced by $g$)
\beq S_{em}=\frac{1}{16\pi}\int d^{4}x\sqrt{g}F_{\mu\nu}F^{\mu\nu}
  \mbox{,\hspace{10mm}}F_{\mu\nu}=\pt_{\mu}A_{\nu}-\pt_{\nu}A_{\mu}
\eeq
\beq S_{higgs}=\frac{1}{4\pi}\int d^4 x\sqrt{g}
  \left[g^{\mu\nu}(\partial_{\mu}+ieNA_{\mu})\vph
  (\partial_{\nu}-ieNA_{\nu})\vph^{*}
  +\frac{\lm}{2}\left(|\vph|^{2}-\frac{v^{2}}{2}\right)^{2}\right]
\eeq
and $S_{mat}$ is the action of a matter field $\xi$ of charge
$Q_{\xi}=\hbar e$ that does not condensate. Consequently, although the
$U(1)$ symmetry is spontaneously broken, there is a residual symmetry
which is respected by the vacuum:
\beq \Om=e^{i2\pi\frac{k}{N}}\in\Zb_{N}\subset U(1) \eeq
under which $\xi\goto \Om\xi$ and $\vph$ and $A_{\mu}$ are invariant.
 
The existence of a $\Zb_{N}$ (primary) hair would mean that the
black-hole states are labeled by a $\Zb_{N}$ charge which can be
measured by an external observer. CPW propose to identify the charge
dependence through the canonical partition function in the black-hole
sector. The partition function for a gauge theory can be expressed by
a Euclidean path integral \cite{GPY}:
\beq Z(\bt)\equiv\mbox{tr}(e^{-\bt H})=
  \int_{\bt\hbar}\Dc A_{\mu}\Dc\phi_{i}e^{-S_{E}/\hbar}
\eeq
where $\phi_{i}$ represent the fields of the model, $S_{E}$ is the
Euclidean action and the integration $\int_{\bt\hbar}$ is over the
configurations which are periodic in the Euclidean time $\tau$ with
period $\bt\hbar$ and satisfy
\beq A_{\tau}(\tau,\vec{x})\goto0
  \mbox{\hspace{5mm} as \hspace{5mm}} |\vec{x}|\goto\infty.
\eeq
As suggested by Gibbons and Hawking \cite{GibHawk}, to treat
gravitation quantum mechanically, one integrates over (Euclidean)
geometries as well. For the black-hole sector the geometries are
restricted to a $\RR^{2}\times S^{2}$ topology.
 
Since the charge $Q$ commutes with the Hamiltonian $H$, it is
meaningful to consider
the partition function $Z(\bt,Q)$, restricted to states of a given
charge $Q$. For $U(1)$ gauge group these are, by definition, states
$|Q>$ which satisfy
\beq U(\Om)|Q>=e^{i\om Q/\hbar e}|Q> \eeq
where $\Om(\vec{x})$ is a gauge transformation with the asymptotic
form
\beq \Om(\vec{x})\goto e^{i\om}=
  \mbox{const. \hspace{5mm} as \hspace{5mm} }|\vec{x}|\goto\infty
\eeq
and $U(\Om)$ is the unitary representation of the gauge group in the
space of states. Inserting the appropriate projection operator $\Pc_Q$
into the trace one obtains
\beq Z(\bt,Q)\equiv\mbox{tr}(\Pc_Q e^{-\bt H})=
  \int_{-\infty}^{\infty}\frac{d\om}{2\pi}
  e^{-i\om\frac{Q}{\hbar e}}\tilde{Z}(\bt,\om)
\eeq
where
\beq \tilde{Z}(\bt,\om)=\int_{\bt\hbar,\om}\Dc g_{\mu\nu}
  \Dc A_{\mu}\Dc \phi_{i}e^{-S_{E}/\hbar}
\eeq
and the integration $\int_{\bt\hbar,\om}$ is over the configurations
which are periodic in $\tau$ with period $\bt\hbar$ and satisfy
\beq \int_{0}^{\bt\hbar}eA_{\tau}d\tau\goto\om
  \mbox{\hspace{5mm} as \hspace{5mm} }|\vec{x}|\goto\infty.
\eeq
The functional integral in $\tilde{Z}(\bt,\om)$ is calculated
semiclassically therefore only configurations with finite action
contribute. This means that the geometry must be
asymptotically flat. In the Higgs model this also means
$|D\vph|\goto0$ and $|\vph|\goto\frac{1}{\sqrt{2}}v$ which implies
\beq \om=2\pi\frac{k}{N}\mbox{ , }k\in\Zb \eeq
($k$ is called ``vorticity''), so the expression for the partition
function is, finally
\beq Z(\bt,Q)=\frac{1}{N}\sum_{k=-\infty}^{\infty}
  e^{-i2\pi\frac{k}{N}\cdot\frac{Q}{\hbar e}}\tilde{Z}(\bt,k)
\eeq
\beq \tilde{Z}(\bt,k)=\int_{\bt\hbar,k}e^{-S_{E}/\hbar}
\eeq
and using standard thermodynamics one obtains a Q-dependent correction
to the relation between the mass of the black-hole and its temperature:
\beq M(\bt,Q)-M(\bt,Q=0)=-\derp{\bt}\ln\frac{Z(\bt,Q)}{Z(\bt,Q=0)}.
\eeq
The same method provides expectation values for local observables such
as the components of the electric field $E_{j}=iF_{j\tau}$,
in a given Q-sector:
\beq <E_{j}>_{\bt,Q}=\frac{1}{Z(\bt,Q)}\frac{1}{N}
  \sum_{k=-\infty}^{\infty}e^{-i2\pi\frac{k}{N}\cdot\frac{Q}{\hbar e}}
  \int_{\bt\hbar,k}iF_{j\tau}e^{-S_E/\hbar}
\eeq
and a non trivial Q-dependence provides (at least conceptually) a
method to measure the $\Zb_{N}$-charge.
 
In the semiclassical limit $\hbar\goto0$, the functional integral is
dominated by the configuration which minimizes the Euclidean action
and therefore is a solution of the classical (Euclidean) equations of
motion -- an instanton:
\beq \tilde{Z}(\bt,k)\sim e^{-S_{clas}/\hbar}. \eeq
There is an instanton solution for each vorticity $k$. The
corresponding action $S_{k}$ is even in $k$ and increases with $|k|$,
therefore the leading $Q$-dependence is obtained by retaining only
$k=-1,0,1$. The result is
\beq \frac{Z(\bt,Q)}{Z(\bt,Q=0)}\approx
  1-2\left[1-\cos\left(\frac{2\pi Q}{N\hbar e}\right)\right]
  \frac{\tilde{Z}(\bt,k=1)}{\tilde{Z}(\bt,k=0)}
\eeq
with
\beq \frac{\tilde{Z}(\bt,k=1)}{\tilde{Z}(\bt,k=0)}
  \approx\frac{1}{2}Ce^{-\Dl S/\hbar}
\eeq
where $\Dl S=S_{1}-S_0$ and $\frac{1}{2}C(\bt\hbar)$ is a ratio of
functional determinants (which can be shown to be positive and
$\Oc(1)$ in the limit $\hbar\goto0$, $\bt\hbar=$const.). This leads
(neglecting terms suppressed by additional powers of $\hbar$) to:
\beq\label{DM} M(\bt,Q)-M(\bt,Q=0)\approx-
  \left[1-\cos\left(\frac{2\pi Q}{N\hbar e}\right)\right]
  C\frac{\pt\Dl S}{\pt\bt\hbar}e^{-\Dl S/\hbar}.
\eeq
The final step is the calculation of $C$ and $\Dl S$. $S_{0}$ is the
action of the (Euclidean) Schwarzschild black-hole, which is (after
substructing the (infinite) flat action)
\beq S_{0}=\frac{(\bt\hbar)^{2}}{16\pi G} \eeq
For $k=1$ one seeks
a rotationally invariant solution that obeys the boundary conditions at
spatial infinity $r\goto\infty$. The corresponding action is evaluated
for two limiting values of the Compton wavelength of the massive
vector meson \mbox{$\mu^{-1}=(Nev)^{-1}$}, called ``string thickness'':
\begin{itemize}
 \item{The thick string limit $\mu^{-1}\gg r_{+}$:}
 
  (where $r_+\approx\frac{\bt\hbar}{4\pi}$ is the radius of the
  horizon)
 
  In this limit the action can be expanded in the small parameter
  \linebreak\mbox{$(\mu r_+)^{2}=(r_+Nev)^{2}$} and the leading term is
  obtained by setting $v^{2}=0$ which corresponds to the unbroken
  $U(1)$ gauge theory. Consequently the solution is the Euclidean
  Reissner-Nordstr\"{o}m black-hole, so $\Dl S$ is given by
  \beq\label{DLSthick}
    \Dl S\approx\frac{\pi}{2(Ne)^{2}}\left[1+\frac{1}{2}
    \frac{G}{(\bt\hbar)^{2}}\left(\frac{2\pi}{Ne}\right)^{2}\right].
  \eeq
 \item{The thin string limit $\mu^{-1}\ll r_{+}$:}
 
  In this limit it is argued that the solution in the $r-\tau$ plane
  looks just like a cross section of a cosmic string at the origin
  $r=0$ (recall that $\tau$ here is an angular coordinate) and using
  an effective action one obtains
  \beq \Dl S\approx\frac{1}{4\pi}(\bt\hbar)^{2}T_{st}(1-4GT_{sr})^{-1}
  \eeq
  where $T_{st}$ is the string tension. Small back reaction
  corresponds to $GT_{st}\ll1$ and in this limit it can be shown that
  $T_{st}$ has the form
  \beq T_{st}\approx\frac{1}{4}v^{2}f(\lm/e^{2}) \eeq
  where $f$ is a slowly varying function such that $f(1)=1$ therefore
  for small back reaction one has
  \beq\label{DLSthin}
    \Dl S\approx\frac{(\bt\hbar)^{2}}{16\pi}v^{2}f(\lm/e^{2}).
  \eeq
\end{itemize}
To summarize, by calculating the canonical partition function in
different $Q$-sectors, one discovers observable quantities (\eg mass,
electric field) which depend on the $\Zb_{N}$ charge $Q$ of the
black-hole and consequently enable a measurement of this charge.
 
\section{The Elusiveness of the Hair}
In the process of getting the results of the previous section, a few
simplifying assumptions were made -- an inevitable step in the
current state of knowledge of quantum gravity. A necessary condition
(though not sufficient) for
the validity of the results is that they be consistent with the
assumptions. The assumption that underlies the whole method (as stated
by the authors themselves) is that to calculate the effects of the
$\Zb_{N}$ charge, one can neglect the Hawking radiation. Indeed, the
use of a canonical partition function applies to a stationary system
in thermal equilibrium with a heat reservoir (in our case this
translates to a black-hole in equilibrium with a surrounding radiation
bath). It is not clear if the very existence of such a system is
allowed by the laws of physics (Since it does not satisfy Einstein's
equations). So in writing down the canonical partition function, one
actually refers to an evaporating black-hole with the hope that it
can be considered in an {\em approximate} equilibrium (a notion that
seems to make sense semiclassically since the evaporation process is
a quantum effect -- of order $\hbar$).
 
Apostriori, the problem with this assumption is that the effect that
was discovered is non-perturbative in $\hbar$ and therefore much
weaker then the neglected Hawking radiation. Let us illustrate this.
We are concerned with the limit $\hbar\goto0$, $\bt\hbar=$const., in
which $\bt$ is very large, and therefore
\beq Z(\bt,Q)\approx<Q|e^{-\bt H}|Q> \eeq
where $|Q>$ is the lowest-energy state with charge $Q$. We define the
``topological vacua''
\beq |n>\equiv\sum_{q=0}^{N-1}e^{-i2\pi\frac{nq}{N}}|Q=q\hbar e> \eeq
and obtain
\beq <n+\hat{k}|e^{-\bt H}|n>=\hat{Z}(\bt,\hat{k})\equiv
  \sum_{N|k-\hat{k}}\tilde{Z}(\bt,k).
\eeq
(Note that $n$ and $\hat{k}$ are defined modulu $N$.) This means that
$\hat{Z}(\bt=\frac{i}{\hbar}t,\hat{k})$ is the transition amplitude
between topological vacua with $\Dl n=\hat{k} \bmod N$, during the
time period $t$:
\beq \hat{Z}(\bt=\frac{i}{\hbar}t,\hat{k})
  =<n+\hat{k}|e^{-\frac{i}{\hbar}tH}|n>
\eeq
(and the leading contribution $\tilde{Z}(\bt,k)$ is the one with the
smallest $|k|$)\footnote{
  The transition (tunneling) between neighboring topological vacua
  $|k|=1$ was interpreted in \cite{CPW} as a process of nucleation
  and annihilation of a virtual cosmic string which envelopes the
  black-hole, however this interpretation is not required for our
  results.}.
This is the exact situation as in gauge field theories (or for a
particle in a one dimensional periodic potential). The existence of
Euclidean configurations labeled by a ``topological charge'' (in the
present case it is the vorticity $k$) which contribute to the vacuum
transition amplitude, implies the existence of a set of states, called
``topological vacua'' between which the Euclidean configurations
interpolate. (This follows from the usual interpretation of Euclidean
configurations as contributions to tunneling amplitudes.)
 
The statement that the states $\{|Q>\}$ are non degenerate is
equivalent to the statement that the topological vacua are not
stationary states, or in other words, the transition amplitudes
between different vacua do not vanish. In the context of a
semiclassical calculation such a statement can be checked only for
times which are not longer then the semiclassical life-time of a
black-hole\footnote{
  Actually, since the Hawking radiation was neglected, the time should
  be shorter then the mean time between the emission of
  Hawking-radiation quanta, which is (for photons with energy
  $\sim\frac{1}{\bt}$)
  \[ t\sim\frac{240}{\pi}\bt\hbar. \] }
which is
\beq\label{t} t\approx \frac{10\hbar^2}{\pi^2 G}\bt^3. \eeq
Let us calculate the transition amplitude:
\begin{eqnarray}
 \Ac_{\Dl n}(t) & \equiv & <n+\Dl n|e^{-\frac{i}{\hbar}tH}|n>= \\
   & = & \sum_{q=0}^{N-1}e^{i2\pi\frac{q\Dl n}{N}}
         <e^{-\frac{i}{\hbar}tH}>_{Q=q\hbar e}. \nonumber
\end{eqnarray}
Using eq. (\ref{DM}) we have
\beq M(q)\equiv<H>_{Q=q\hbar e}\approx
  M_0+2\Dl M\cos\left(2\pi\frac{q}{N}\right)
\eeq
where
\beq 2\Dl M=C\frac{\pt\Dl S}{\pt\bt\hbar}e^{-\Dl S/\hbar} \eeq
therefore
\begin{eqnarray}
 \Ac_{\Dl n}(t) & = & e^{-\frac{i}{\hbar}tM_0}\sum_{q=0}^{N-1}
   e^{i2\pi\frac{q\Dl n}{N}}\sum_{l=0}^{\infty}\frac{1}{l!}
   \left[-\frac{i}{\hbar}t\Dl M\left(e^{i2\pi\frac{q}{N}}+
   e^{-i2\pi\frac{q}{N}}\right)\right]^l  \nonumber \\
 & = & Ne^{-\frac{i}{\hbar}tM_0}\sum_{l_+,l_-}\frac{1}{l_+!l_-!}
   \left(-\frac{i}{\hbar}t\Dl M\right)^{l_++l_-}
\end{eqnarray}
where the sum is over all non-negative integers $l_+,l_-$ which
satisfy
\beq l_+-l_-=\Dl n\bmod N. \eeq
Without loss of generality, we choose $|\Dl n|\leq\frac{N}{2}$ so for
$|t\Dl M|\ll\hbar$ we have\footnote{
  Actually for $|\Dl n|=\frac{N}{2}$ the leading term is twice bigger
  but this does not spoil our argument}
\beq |\Ac_{\Dl n}(t)|=\frac{N}{\Dl n!}\left|\frac{t\Dl M}{\hbar}
  \right|^{|\Dl n|}\left(1+\Oc\left(\frac{t\Dl M}{\hbar}\right)\right)
  \ll1.
\eeq
Finally we evaluate $t\Dl M/\hbar$ taking $\Dl S$ from eqs.\
(\ref{DLSthick} -- \ref{DLSthin}) and $t$ from (\ref{t}):
\begin{itemize}
 \item for the thick string
  \beq \left|\frac{t\Dl M}{\hbar}\right|\stackrel{<}{\sim}
    \frac{40}{\pi}C\left(\frac{\Dl S}{\hbar}\right)^2 e^{-\Dl S/\hbar}
  \eeq
  with
  \beq \frac{\Dl S}{\hbar}\stackrel{>}{\sim}\frac{\pi}{2\al} \eeq
  where
  \beq \al\equiv\hbar(Ne)^2 \eeq
  is the dimensionless gauge field coupling constant (analogous to the
  fine structure constant in electromagnetism).
 \item for the thin string (neglecting back reaction)
  \beq \left|\frac{t\Dl M}{\hbar}\right|\approx\frac{40}{\pi}C
    \frac{1}{GT_{st}}\left(\frac{\Dl S}{\hbar}\right)^2 e^{-\Dl S/\hbar}
  \eeq
  with
  \beq \frac{\Dl S}{\hbar}\approx\frac{(\bt\hbar)^2}{16\pi\hbar}v^2
    f\left(\frac{\lm}{e^2}\right)\gg\frac{1}{\al}
  \eeq
  (the inequality follows from
  $\frac{\bt\hbar}{4\pi}\approx r_+\gg(Nev)^{-1}$).
\end{itemize}
Both expressions vanish exponentially with $1/\al$ and since $\al$
must be small, it follows that indeed $|t\Dl M|/\hbar$ is negligibly
small.
 
To summarize, we have shown that the semiclassical approximation
does not give any reliable indication for the instability of the
topological vacua -- the transition between such states during the
semiclassical life of the black-hole is extremely improbable. This
means that there is no indication for the non-degeneracy of the $|Q>$
states -- no $\Zb_N$-hair. This, of course, does not prove that the
quantum hair does not exist and to discover what really happens, one
must go beyond the semiclassical approximation. Since we do not know
how to do that, it would be useful to get intuition about what can
be expected, by analyzing a simpler model which shares with the real
problem the relevant properties. We will analyze such a toy model in
the next section, concluding that the $\Zb_N$ hair is indeed
expected to be illusory.
 
\section{A Toy Model}\label{Model}
\begin{figure}
 \begin{center} \leavevmode \epsfbox{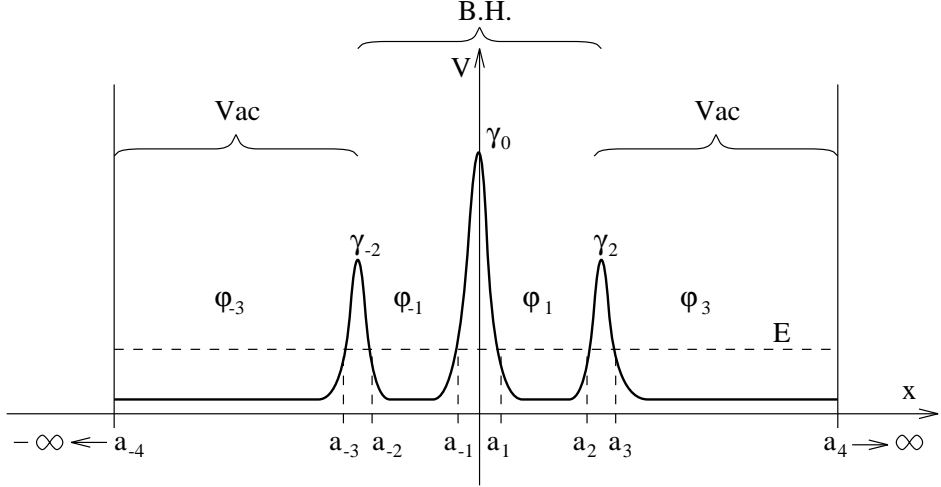} \end{center}
 \caption{The Potential}\label{Pot}
\end{figure}

We will consider a particle in a one dimensional space, subjected to a
potential as described in figure \ref{Pot}.
There are three high potential-barriers (labeled $-2,0,2$) which
divide the space into four regions (labeled $-3,-1,1,3$). The two
central regions represent the black-hole sector and the barrier
between them is what causes the splitting of the spectrum of this
sector -- the analogue of the quantum hair. The outer regions represent
the ``vacuum'' into which the black-hole evaporates
and the Hawking radiation is modeled by the penetration through
the side barriers. This is of course a very crude model and it is only
meant to demonstrate how the instability of the internal states
influences the ability to measure the splitting. The details of the
analysis are given in the appendix and here we only summarize the
results.
 
The stationary state of energy $E$ can be written in each region,
in the WKB approximation (assumed to be valid there, except near the
sides), in the form:
\begin{itemize}
 \item Between the barriers ($V(x)<E$):
  \beq
   \Psi_{l}=\frac{1}{\sqrt{k}} \left[
   A_{l}^{+}e^{i\left|\int_{a_{l}}^{x}kdx'\right|}+
   A_{l}^{-}e^{-i\left|\int_{a_{l}}^{x}kdx'\right|}\right]
  \eeq
 \item ``Under'' the barriers ($V(x)>E$):
  \beq
   \Psi_{l}=e^{-i\frac{\pi}{4}}\frac{1}{\sqrt{2k}} \left[
   2B_{l}^{+}e^{\left|\int_{a_{l}}^{x}kdx'\right|}+
   B_{l}^{-}e^{-\left|\int_{a_{l}}^{x}kdx'\right|}\right]
  \eeq
 \item where $\hbar k(x)=\sqrt{2m|E-V(x)|}$.
\end{itemize}
The relations between the parameters in adjacent regions are given
by the ``connection formula'' for each ``turning point''
($V(x)=E$)\footnote{
  Note that each region is described by two expressions, one for each
  side of the region. This and the unusual normalization of the
  $B$-coefficients are meant to make these relations as simple as
  possible (see the appendix).}.
All the characteristics of the system can be described by the
following parameters (functions of the energy $E$):
\begin{description}
 \item A barrier is represented by the ``penetrability''
  \beq \gm=\frac{1}{2}e^{-\int kdx} \eeq
 \item and an allowed region (between barriers) is represented by
  the ``total phase''
  \beq \vph=\int kdx \eeq
 \item (where the integration is over the relevant region)\footnote{
    For the validity of the connection formula, we must have
    \[ \gm\ll 1 \mbox{\hspace{5mm} and \hspace{5mm}}
       \vph\stackrel{>}{\sim} 10. \]
    The first restriction is unimportant since anyway we are interested
    only in barriers very high above the energy level. But the second
    restriction means that the analysis is invalid for the lowest levels.
    This restriction does not appear in the alternative model in which
    the turning points are replaced by vertical walls. At first look
    the results relevant to our purposes seem to be identical in both
    models.}.
\end{description}
We will compare probabilities corresponding to different
wave-functions (see below) and for this they have to be normalized.
Therefore we will perform the
quantization in a box, very large compared to any relevant length
parameter in the problem.
 
For $\gm_{0,\pm1}\goto0$ (infinite barriers) there are four
disconnected regions and each region has its own states. The inner
(``black-hole'') energy spectrum is determined by the quantization
conditions
\beq \cot\vph_{\pm1}=0. \eeq
We will be interested in a degenerate level, \ie we will consider a
neighborhood of an energy $E^{0}$ which is in the spectrum of both
left and right inner regions:
\[ \cot\vph_{1}^{0}=\cot\vph_{-1}^{0}=0. \]
For this level we will check when it is possible to recognize a
splitting caused by finiteness of the central barrier ($\gm_0>0$).
 
\subsection{The Symmetric Case}
In analogy to the $\Zb_N$ symmetry of the black-hole we first assume
a $\Zb_2$ symmetry of the potential:
\beq\label{Sym0} V(-x)=V(x). \eeq
Consequently, the solutions are either symmetric or antisymmetric.
Defining
\beq \bt_1\equiv B_1^+/B_1^- \eeq
it can be shown that a symmetric solution corresponds to
$\bt_{1}=+\gm_{0}$ and an antisymmetric one corresponds to
$\bt_{1}=-\gm_{0}$ therefore $\bt_{1}/\gm_{0}=\pm1$ should be
identified as the $\Zb_2$-charge of a state.
 
We want to explore the conditions in which the charge and
the energy of the black-hole states are correlated in an observable
way so that the charge would be a genuine (\ie measurable) quantum
hair. In the limit $\gm_{\pm 2}\goto 0$, which corresponds to a
completely stable black-hole, the correlation is clear. The black-hole
sector is isolated and every degenerate level splits to two levels
\beq dE\equiv E-E^0=\pm\frac{\gm_{0}}{\pi D_1} \eeq
where
\beq D\equiv\frac{dn}{dE}=\frac{1}{\pi}\frac{d\vph}{dE} \eeq
is the (local) density of states. This splitting corresponds to
\beq d\vph_1\equiv\vph_1-\vph_1^0=\pm\gm_0=-\bt_1 \eeq
and this means that we have a separation between states of different
charge. Note also that the separation is proportional to the
``penetrability'' $\gm_0$ of the barrier.
 
When the side barriers are finite (\ie the black hole is not stable),
the states are not restricted to a single region. In particular, there
are no exclusively-black-hole states (each state has ``tails'' in the
outer regions) so we must investigate the whole spectrum of states.
This is a very dense spectrum (tends to the continuum in the infinite
box limit) and both charges are equally distributed in it, so
correlations can emerge only from differences in the probabilities to
find (in the black-hole region) states of different charges in
different regions of energy. In the appendix we calculate the
probability $\Pc_1$ to find the particle in the right black-hole
region as a function of charge (represented by $\bt_1$) and energy
(represented by $d\vph$). The results are shown in figure \ref{P1sym}:
\begin{figure}
 \begin{center} \leavevmode \epsfbox{fig2.ps} \end{center}
 \caption{$\Pc_1$ as a function of $d\vph_1$ for the symmetric case}
 \label{P1sym}
\end{figure}
for each charge\ldots
\begin{enumerate}
 \item the maximal $\Pc_{1}$ is obtained at
   \beq d\vph_{1}=-\bt_{1}=\pm\gm_{0}; \eeq
 \item the ``width'' $2\Dl\vph_{1}$ at half height is
   \beq \Dl\vph_{1}\approx\gm_{2}^{2}; \eeq
\end{enumerate}
so each level in the $\gm_2\goto0$ limit is replaced by a band of
levels with width $\gm_2^2$. This is actually a recovery of a very
well-known phenomenon -- the width of a resonance.
 
Now the situation should be clear. When the instability $\gm_2^2$ of
the black hole is small compared to the penetrability $\gm_0$ through
the central barrier we have a sharp separation between
the two bands of opposite charge and the charge is therefore
measurable (this is the situation, in particular, for $\gm_{2}=0$).
As the instability increases, the bands get wider and when
$\gm_{2}^{2}\gg\gm_{0}$ they overlap completely and there is no
practical way to determine the charge knowing the energy:
the instability of the internal states cancels the splitting, as
anticipated.
 
\subsection{The General Case}
Strictly speaking, the symmetric potential is the suitable model for
the real situation, so it should be explained why it is interesting to
consider a more general case. One motivation is to see if the results
change by a small perturbation (\ie in the case of an approximate
symmetry). Also, since this model is very naive, it
will be reassuring to realize that the desired effect is more general
and exists also when the degeneracy is accidental
We indeed expect it to be universal because the phenomenon of a width
of a resonance is commonly seen as a consequence of the uncertainty
principle
\[ \Dl E\Dl t\stackrel{>}{\sim}\hbar. \]
 
As in the symmetric case, our task is to identify, in the
neighborhood of a degenerate energy, two bands of states and to
determine their width and the distance between them. One might
expect that the qualitative situation is the same as in the symmetric
case (especially when there is an approximate symmetry), namely that
there are two (approximate) values of $\bt_{1}$ and to each of them
corresponds a band in a different location. This expectation was found
to be wrong, but the conclusions nevertheless remains unchanged.
As shown in the appendix, assuming a small instability $\gm_{\pm2}^2$,
the quantization condition leads this time to pairs of states, one
with $\bt_1\approx\cot\vph_1$ and the other with
$\bt_1\approx\gm_0^2/\cot\vph_{-1}$. As illustrated in figure
(\ref{P1gen})
\begin{figure}
 \begin{center} \leavevmode \epsfbox{fig3.ps} \end{center}
 \caption{$\Pc_{1}$ as a function of $d\vph_{1}$ for the general case}
 \label{P1gen}
\end{figure}
for each of these cases, $\Pc_1$ is an even function of $d\vph_{1}$
with two local maxima around $d\vph_1=\pm\gm_0/\lm_0$, where
\beq \lm_0^2\equiv\left.\frac{d\vph_{-1}}{d\vph_1}\right|_{E^0} \eeq
($\lm_0^2$ can be recognized as the ratio of the (local) densities of
states in the two black-hole regions). The only difference between
these cases is the width of the minima but in both cases it is
proportional to the instability $\gm_{\pm2}^2$ of the black hole.
 
Now we see that the results of the symmetric case are in fact general:
a degenerate level splits into two bands which are well separated for
a relatively stable black hole $\gm_{\pm2}^{2}\ll\gm_{0}$. The width
is proportional to the instability $\gm_{\pm2}^{2}$ so as
$\gm_{\pm2}^{2}$ increases the band gets wider and for
$\gm_{\pm2}^{2}\gg\gm_{0}$ one should expect that there will be no
trace of the splitting.
 
\begin{figure}
 \begin{center} \leavevmode \epsfbox{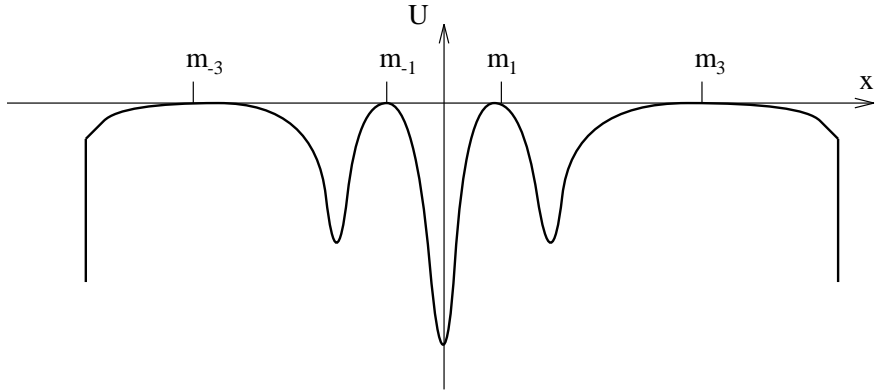} \end{center}
 \caption{The Euclidean potential $U=-V$}
 \label{InvPot}
\end{figure}

Finally, to see in what way the thermodynamical analysis of the black
hole fails, let us apply the same method to this quantum
mechanical model. For the sake of simplicity, we assume
$V_{\mbox{min}}=0$ and the minimum is reached only once in each region
(at $m_{-3},m_{-1},m_1,m_3$ respectively). The path-integral
expression for the partition function is
\beq Z(\bt)=\int_{\bt\hbar}\Dc[x(\tau)]e^{S_E/\hbar} \eeq
where the integration is over paths $x(\tau)$ periodic in the
Euclidean time $\tau$ with period $\bt\hbar$. In the semiclassical
limit $\hbar\goto0$ the integral is dominated by classical paths of
a particle in an inversed potential $U=-V$ (see figure \ref{InvPot})
and in the limit $\tau\goto\infty$ only paths between minima
contribute. These limits correspond to $\bt\goto\infty$ (low
temperature) which means that only the lowest-energy states contribute
to the partition function.
 
When the barriers are infinite ($\gm_{0,\pm2}\goto0$), the only finite
action paths are constant paths and these give the degenerate
ground-state energy at the two black-hole regions. When $\gm_0$ is
finite there are additional paths going between the minima $m_{-1}$
and $m_1$ and their contribution is what removes the degeneracy. These
are the instantons, analogous to those used in the CPW model. But when
$\gm_{\pm2}$ are also finite, there are still
additional paths going to and from the vacuum areas. As we saw, these
paths change the spectrum significantly (unless $\gm_{\pm2}$ are
sufficiently small). This suggests that to correct the calculation of
the black hole energy spectrum, one should include configurations
analogous to this last type of paths. These might be configurations
that interpolate between (Euclidean) black holes of different mass
(since we know that these are connected through the process of
Hawking radiation).

\section{General Consequences}
In this work we have shown that the quantum hair suggested by CPW is
most probably illusory. Essentially, this follows directly from the
fact that the effect was much smaller than the Hawking radiation
therefore the same arguments should apply to a larger class of
would-be quantum hair, in particular those that correspond to
non-perturbative effects, as the one analyzed here. This puts the
search for quantum hair in a rather delicate situation. On the one
hand, the classical theory implies that these hair must be weak --
vanishing in the $\hbar\goto0$ limit. On the other hand, the
instability of the black hole implies that they cannot be too weak
otherwise they will be shadowed by the evaporation process. Another
aspect of this problem (not unrelated) is the fact that so far we only
know how to make semiclassical calculations and these are considered
reliable only for large black holes (small curvature) and weak effects
-- typically too weak in view of the above arguments. All this leads
to a conclusion that there is little hope to find genuine quantum hair
using semiclassical methods. Perhaps these arguments can be made more
precise to form a quantum extension of the classical ``no hair''
theorems.
 
Beyond the semiclassical approximation the situation may be quite
different. In particular, it is possible that the black hole does not
evaporate completely and instead stabilizes as a Planck-scale remnant
\cite{ACN}. Our arguments do not apply to this case and such a remnant
might have arbitrarily weak quantum hair. Thus until we understand
Planck-scale physics, quantum hair cannot be excluded. This however
should not make a difference for large black holes which seem to be
bald also quantum-mechanically.
 
\section*{Appendix: Calculational details}
In this appendix we derive the results of section \ref{Model}.
The quantization in a box implies -- by
conservation of probability -- that the probability current vanishes
and this implies (With the notation defined in section \ref{Model}):
\beq |A_{l}^{+}|=|A_{l}^{-}| \eeq
and
\beq\label{def-al}
 \bt_{l}\equiv \frac{B_{l}^{+}}{B_{l}^{-}}\mbox{ is real.}\eeq
The first of these relations means that
$|A_{l}|\equiv|A_{l}^{+}|=|A_{l}^{-}|$ is a suitable measure for the
probability density. More precisely, the probability for finding the
particle in a certain allowed region with a side $a_l$ is
\beq\label{Pdef}
  \Pc_{l}\approx\pi\frac{\hbar^{2}}{2m}D_{l}|A_{l}|^{2}.
\eeq
where $D_l$ is the (local) density of states\footnote{
  This follows from direct differentiation
  \[ D\approx \frac{m}{\pi\hbar^{2}}\frac{\Dl x}{\bar{k}} \]
  where $\bar{k}$ is the average wave number:
  \[ \frac{1}{\bar{k}}\equiv\frac{1}{\Dl x}\int\frac{dx}{k}\mbox{ .}
  \]
  Note also that $D$ is proportional to the size $\Dl x$ of the region
  so in the external regions the density is very large (this is
  extensively used in the following).}:
\beq
 D\equiv\frac{dn}{dE}=\frac{1}{\pi}\frac{d\vph}{dE}.
\eeq
The relations among the amplitudes are determined by the connection
formula, which obtains in the present notation the simple form
\beq \left(\begin{array}{c} B^+ \\ B^- \end{array}\right)=
  \frac{1}{\sqrt{2}}\left(\begin{array}{cc} 1 & i \\ i & 1 \end{array}
  \right)\left(\begin{array}{c} A^+ \\ A^- \end{array}\right).
\eeq
The relations between the $B$'s across a barrier are
\beq \left(\begin{array}{c} B_q^- \\ B_q^+ \end{array}\right)=
  \left(\begin{array}{cc} \gm^{-1} & 0 \\ 0 & \gm \end{array}\right)
  \left(\begin{array}{c} B_p^+ \\ B_p^- \end{array}\right)
\eeq
and the relations across an allowed region are
\beq \left(\begin{array}{c} A_q^+ \\ A_q^- \end{array}\right)=\left(
  \begin{array}{cc} e^{-i\vph} & 0 \\ 0 & e^{i\vph} \end{array}\right)
  \left(\begin{array}{c} A_p^- \\ A_p^+ \end{array}\right)
\eeq
which implies
\beq \left(\begin{array}{c} B_q^+ \\ B_q^- \end{array}\right)=
  \left(\begin{array}{cc} \cos\vph & -\sin\vph
  \\ \sin\vph & \cos\vph \end{array}\right)
  \left(\begin{array}{c} B_p^- \\ B_p^+ \end{array}\right)
\eeq
($a_p,a_q$ denote the sides of the relevant region).
From these relations follow some simple and important properties of
the real parameter $\bt_{l}$ defined in (\ref{def-al}) which make it a
central parameter in the analysis:
\begin{itemize}
 \item it is directly connected to the ratio of amplitudes on both
   sides ($a_{p}$, $a_{q}$) of a barrier:
   \beq\label{AtoA}
     \left| \frac{A_{q}}{A_{p}}\right|^{2} =
     \frac{\gm^{-2}\bt_{p}^{2}+\gm^{2}}{\bt_{p}^{2}+1}
   \eeq
   (this is a monotonically increasing function of $\bt_{p}$, going
   from $\gm^{2}$ to $\gm^{-2}$);
 \item  there are simple relations among the $\bt$'s:
   \begin{eqnarray}
     \mbox{over a barrier:} & & \bt_{p}\bt_{q}=\gm^{2} \\
     \mbox{over an allowed region:}\label{allow} & &
     \frac{\bt_{p}+\bt_{q}}{1-\bt_{p}\bt_{q}}
     =\kp\equiv\cot\vph \label{al2}
   \end{eqnarray}
 \item the boundary conditions $\psi(a_{\pm 4})=0$ imply
   \beq \bt_{\pm 3}=\kp_{\pm 3} \eeq
   where
   \beq  \kp_{\pm3}\equiv\cot(\vph_{\pm3}-\frac{\pi}{4}). \eeq
\end{itemize}
Combining all these relations, the quantization condition can be
expressed as
\beq\label{q1} \bt_{-1}\bt_{1}=\gm_0^2 \eeq
where
\beq\label{q2}
  \bt_{\pm 1}=\frac{\kp_{\pm 1}-\frac{\gm_{\pm 2}^{2}}{\kp_{\pm 3}}}%
              {1+\kp_{\pm 1}\frac{\gm_{\pm 2}^{2}}{\kp_{\pm 3}}}.
\eeq
 
\paragraph{The symmetric case:} The $\Zb_2$-symmetry (\ref{Sym0}) of
the potential implies\footnote{
  Note that the following analysis does not use directly the symmetry
  (\ref{Sym0}) of the potential and it is only needed that
  (\ref{Sym1}) will be satisfied in a neighborhood of $E_{0}$.}
\beq\label{Sym1}
  \kp_{-1}=\kp_{1}\mbox{,\hspace{10mm}}\gm_{-2}=\gm_{2}
  \mbox{,\hspace{10mm}}\kp_{-3}=\kp_{3}
\eeq
and therefore
\beq \bt_{-1}=\bt_{1}=\pm\gm_{0}. \eeq
We assume throughout $\gm\ll1$. Among other things this allows us to
neglect the probability to find the particle under a barrier, so we
have
\beq 2\Pc_{1}+2\Pc_{3}\approx 1. \eeq
Substituting (\ref{Pdef}) and using (\ref{AtoA} -- \ref{allow}) we
obtain
\beq
  \frac{1}{2\Pc_{1}}\approx
  1+\frac{D_{3}}{D_{1}}\left|\frac{A_{3}}{A_{1}}\right|^2\approx
  1+\lm_{2}^{2}[\gm_{2}^{2}+\gm_{2}^{-2}\sin^2(d\vph_{1}+\bt_{1})]
\eeq
where $\lm_{2}=\sqrt{\frac{D_{3}}{D_{1}}}\gg1.$
We recall that for each charge $\bt_{1}$ is constant and from this
follows the shape of $\Pc_1$ as described in section \ref{Model}.
 
\begin{figure}
 \begin{center} \leavevmode \epsfbox{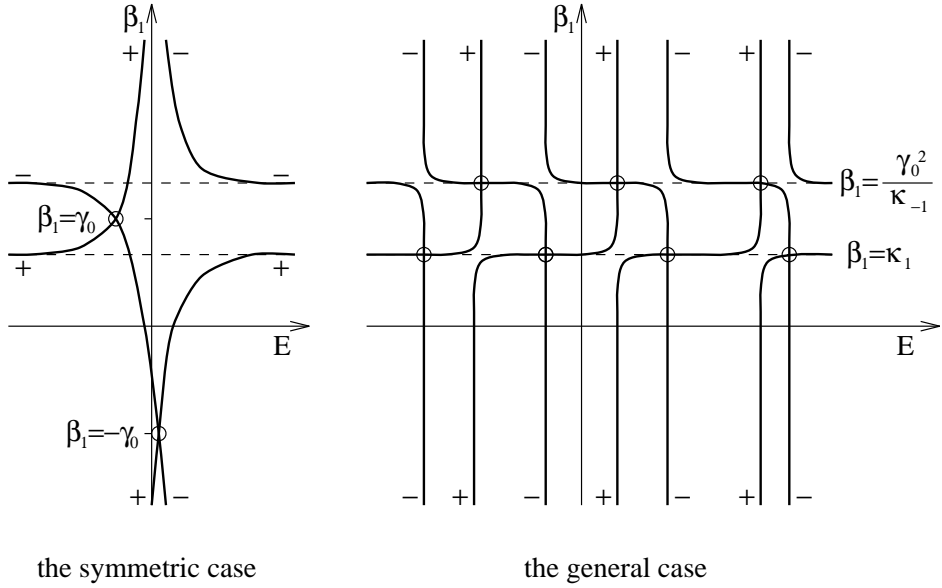} \end{center}
 \caption{The Quantization condition for $\gm_{\pm2}\ll1$}
 \label{Quan}
\end{figure}

\paragraph{The general case:} Proceeding as before, we have
\beq \Pc_{-3}+\Pc_{-1}+\Pc_{1}+\Pc_{3}\approx1 \eeq
which leads to
\begin{eqnarray}\nonumber
  \frac{1}{\Pc_{1}} & \approx & 1+
  \frac{D_{3}}{D_{1}}\left|\frac{A_{3}}{A_{1}}\right|^2
  +\frac{D_{-1}}{D_{1}}\left|\frac{A_{-1}}{A_{1}}\right|^2
  \left(1+\frac{D_{-3}}{D_{-1}}
  \left|\frac{A_{-3}}{A_{-1}}\right|^2\right) \\
  & \approx & 1+\lm_{2}^{2}[\gm_{2}^{2}+\gm_{2}^{-2}\cos^{2}
        (\vph_{1}+\dl\vph_{1})]+   \label{P1} \\
  &   & +\left[\frac{1+\bt_{-1}^{2}}{1+\bt_{1}^{2}}
         \left(\frac{\lm_{0}}{\gm_{0}}\bt_{1}\right)^{2}\right]
        \{1+\lm_{-2}^{2}[\gm_{-2}^{2}+\gm_{-2}^{-2}\cos^{2}
        (\vph_{-1}+\dl\vph_{-1})]\} \nonumber
\end{eqnarray}
where we define:
\beq
  \lm_{0}\equiv\sqrt{\frac{D_{-1}}{D_{1}}}\mbox{ ,\hspace{10mm}}
  \lm_{\pm2}\equiv\sqrt{\frac{D_{\pm3}}{D_{\pm1}}}
\eeq
and
\beq \dl\vph_{\pm1}\equiv\arctan\bt_{\pm1}. \eeq
In the following we assume
\beq
  \lm_{\pm2}^{-2}\ll\gm_{\pm2}^{2}\ll
  \lm_{0}\gm_{0},\frac{\gm_{0}}{\lm_{0}}\ll1
\eeq
\begin{itemize}
 \item $\gm_{0}\ll\lm_{0}\ll\gm_{0}^{-1}$ to insure
   $|d\vph_{\pm1}|\ll1$ in the relevant area;
 \item $\gm_{\pm2}^{2}\ll\lm_{0}\gm_{0},\gm_{0}/\lm_{0}$
   to obtain a sharp splitting;
 \item $\lm_{\pm2}^{-1}\ll\gm_{\pm2}$ to insure sufficient density
   of states.
\end{itemize}
It is reasonable to take for the $\gm$'s and $\lm$'s their values at
$E^{0}$ so, recalling that
\beq\label{df}
  d\vph_{-1}=\lm_{0}^{2}d\vph_{1}\mbox{,\hspace{10mm}}
  \bt_{-1}=\frac{\gm_{0}^{2}}{\bt_{1}}
\eeq
we see that $\Pc_{1}$ can be considered as a function of the two
variables $d\vph_{1}\equiv\vph_{1}-\vph_{1}^{0}$ and $\bt_{1}$.
Therefore to
obtain the function $\Pc_{1}(d\vph_{1})$, we have to find the
dependence of $\bt_{1}$ on $d\vph_{1}$. This is obtained from the
quantization condition (\ref{q1}) and (\ref{q2}). The quantization
points are determined by the intersection of
\beq
  \bt_{1}=\frac{\kp_{1}-\frac{\gm_{2}^{2}}{\kp_{3}}}%
              {1+\kp_{1}\frac{\gm_{2}^{2}}{\kp_{3}}}
\eeq
with
\beq
  \bt_{1}=\frac{\gm_{0}^{2}}{\bt_{-1}}=\gm_{0}^{2}\left[
          \frac{1+\kp_{-1}\frac{\gm_{-2}^{2}}{\kp_{-3}}}%
               {\kp_{-1}-\frac{\gm_{-2}^{2}}{\kp_{-3}}}\right]
\eeq
(see figure \ref{Quan}). It is clear from the graph that
generically we have pairs of states, one with $\bt_{1}\approx\kp_{1}$
(which corresponds roughly to $\kp_{3}\approx0$) and the other with
$\bt_{1}\approx\frac{\gm_{0}^{2}}{\kp_{-1}}$ (which corresponds
roughly to $\kp_{-3}\approx0$)\footnote{
  Other values occur only when $\kp_{3}$ and $\kp_{-3}$ vanish
  (almost) simultaneously and this is very rare if $\vph_{3}$ and
  $\vph_{-3}$ are not correlated. In this respect, the symmetric
  case is exceptional (recall that we get there $\bt_{1}$ independent
  of $\kp_{1}$) and an approximate symmetry is closer to the general
  case then to the symmetric one.}.
We check first the $\bt_{1}\approx\kp_{1}$ states. For
$|d\vph_{1}|\ll1$ we have
\beq \dl\vph_{1}\approx\bt_{1}\approx\kp_{1}\approx-d\vph_{1}. \eeq
Substituting all this to (\ref{P1}) we obtain
\beq
  \Pc_{1}^{-1}\approx(\lm_{2}\gm_{2})^{2}+
  \left(\frac{\lm_{0}}{\gm_{0}}\lm_{-2}\right)^{2}(1+\bt_{-1}^{2})
  \bt_{1}^{2}[\gm_{-2}^{2}+\gm_{-2}^{-2}
  \cos^{2}(\vph_{-1}+\dl\vph_{-1})].
\eeq
This is an even function of $d\vph_{1}$ (recall (\ref{df})), with two
local maxima around \linebreak\mbox{$d\vph_{1}=\pm\gm_{0}/\lm_{0}$} 
(see figure \ref{P1gen}). In the regions of the the maxima
$|d\vph_{1}|\gg\gm_{0}^{2}$ therefore \mbox{$|\bt_{-1}|\ll1$} and we
finally obtain, for the $\bt_{1}\approx\kp_{1}$ states:
\begin{itemize}
 \item the location of the maxima is\footnote{
     Actually the exact location differs slightly from
     $\frac{\gm_{0}}{\lm_{0}}$ but the difference is negligible with
     respect to the width of the band.}:
   \beq d\vph_{1}\approx\pm\frac{\gm_{0}}{\lm_{0}}; \eeq
 \item the maximal value is:
   \beq
     \Pc_{1_{max}}\approx
     [(\lm_{2}\gm_{2})^{2}+(\lm_{-2}\gm_{-2})^{2}]^{-1};
   \eeq
 \item the width $2\Dl\vph_{1}$ at half height is:
   \beq
     \Dl\vph_{1}\approx\frac{1}{2}
     \left(\frac{\gm_{-2}}{\lm_{0}}\right)^2
     \sqrt{1+\left(\frac{\lm_{2}\gm_{2}}{\lm_{-2}\gm_{-2}}\right)^{2}}.
   \eeq
\end{itemize}
The other set of states $\bt_{1}\approx\frac{\gm_{0}^{2}}{\kp_{-1}}$
corresponds to $\bt_{-1}\approx\kp_{-1}$ which is the same as the
first set but for the left black-hole region. For these states
$\Pc_{1}$ looks the same: it has the same maximal values and at the
same locations. Only the width is different:
\beq
  \Dl\vph_{1}\approx\frac{1}{2}(\lm_{0}\gm_{2})^{2}
  \left(\frac{\lm_{-2}}{\lm_{2}}\right)^2
  \sqrt{1+\left(\frac{\lm_{2}\gm_{2}}{\lm_{-2}\gm_{-2}}\right)^{2}}
\eeq
but it is still proportional to $\gm_{\pm2}^{2}$.

\end{document}